\documentclass{mem}
\usepackage{natbib}\usepackage{txfonts}\usepackage{balance}
\usepackage{graphicx}
\usepackage[a4paper]{hyperref}
\idline{75}{282}
\begin{document}
\def\teff{$T\rm_{eff }$}
\def\kms{$\mathrm {km s}^{-1}$}

\title{
Jets and multi-phase turbulence
}

   \subtitle{}

\author{
Martin \,G. \,H. \,Krause\inst{1,2,3}
          }

  \offprints{M. Krause}

\institute{
Max-Planck-Institut f\"ur extraterrestrische Physik,
Giessenbachstrasse,
85748 Garching,
Germany
\and
 Universit\"ats-Sternwarte M\"unchen, 
Scheinerstr. 1, 81679 M\"unchen, Germany
\email{krause@mpe.mpg.de}
\and 
University of Cambridge,
Cavendish Laboratory, Astrophysics Group,
JJ~Thomson Avenue, Cambridge CB3 0HE, UK
}

\authorrunning{Krause}

\titlerunning{Jets and multi-phase turbulence}

\abstract{
Jets are observed to stir up multi-phase turbulence in the inter-stellar 
medium as well as far beyond the host galaxy. Here we present detailed 
simulations of this process. We evolve the hydrodynamics equations with
optically thin cooling for a 3D Kelvin Helmholtz setup with one initial 
cold cloud. The cloud is quickly disrupted, but the fragments remain cold
and are spread throughout our simulation box. A scale free isotropic
Kolmogorov power spectrum is built up first on the large scales, 
and reaches almost down to the grid scale after the simulation time of 
ten million years. \\
We find a pronounced peak in the temperature distribution at 14,000K.
The luminosity of the gas in this peak is correlated with the 
energy. We interpret this as a realisation of the shock ionisation scenario.
The interplay between shock heating and radiative cooling establishes the
equilibrium temperature. This is close to the observed emission in some
Narrow Line Regions.´
We also confirm the shift of the phase equilibrium,
i.e. a lower (higher) level of turbulence produces a higher (lower)
abundance of cold gas. The effect could plausibly lead to a high level of 
cold gas condensation in the cocoons of extragalactic jets, explaining the
so called {\em Alignment Effect}. 

}
\maketitle{}

\section{Introduction}
Jet cocoons are a mixing region, where the subsonic radio synchrotron
emitting jet plasma interacts with the ambient gas.
X-ray \citep[e.g.][]{Sea01} as well as line emitting gas 
\citep[e.g.][also compare Rosario, this proceedings]{BKB04,Nesea07}
is observed cospatial with the radio emission.
Observations clearly show that the line emitting gas as well as the
neutral hydrogen is stirred up and accelerated by the 
highly turbulent jet cocoon plasma
\citep{Morgea2005a}, sometimes also by direct hits of the jet beams. 
The typical velocities reach from a few 
100~km/s in the Narrow Line Region (NLR) of Seyfert galaxies to
more than 1000~km/s in high redshift radio galaxies.
These velocities are much greater than the sound speed in the 
line-emitting and neutral gas, and therefore drive shock waves 
into these components. As the density of the clouds is typically 
of order 100~cm$^{-3}$, the cooling time is short, and the shocks 
will be fully radiative, effectively dissipating the system energy.
This will be observable in sources where the level of photo-ionising
radiation is low, but the emission from radiative shocks will be 
outshone by photo-ionisation, if it is strong enough.
This has indeed been confirmed by line ratio analysis based on 
one-dimensional shock models that include detailed atomic physics
\citep{DS96a}. 

Two-dimensional shock models have shown that 
shocked clouds are torn apart \citep{MKR02}, suppressing star
formation
in this phase \citep{AS08}, but the fragments remain cold.
The kinetic energy is efficiently channelled into
radiation at an enhanced level compared to one-dimensional models,
and turbulence
\citep{SBD03}. In fact, the whole jet cocoon is a turbulent region,
and that paradigm was applied in the two-dimensional simulations by 
\citet[][KA07]{KA07}. They showed that the Kolmogorov power spectrum
establishes from large scales first and  that the kinematics is governed 
by a particular relation between the Mach number and the density.
They also showed that the gas preferentially assembles at a
temperature around 14,000~K, due to an equilibrium of shock heating and 
radiative cooling. Another finding was the rapid growth of the cold
phase. 

Here, we pick up on the simulations of KA07, going to 
three dimensions and varying the ambient gas temperature.

\section{Code and setup}
We use the 3D magnetohydrodynamics code {\em Nirvana} 
\citep{ZY97,mypap01a}, employing a cooling function
for solar metallicity gas, that goes down to arbitrarily low
temperatures (KA07 for details). We keep the setup almost
identical to KA07: a Kelvin-Helmholtz instability with a
density ratio of 10,000 and uniform pressure. 
The density of the denser part is kept fixed at
$10^{-2}$~m$_\mathrm{p}/\mathrm{cm}^3$, and its temperature is varied 
between 1~and~15 Mio~K in different runs. 
The relative Mach number
is 0.8 (80) with respect to the sound speed in the lower (higher)
density gas. Embedded in the denser medium is a still denser cloud
(1, 10~m$_\mathrm{p}$/cm$^3$), also set up in pressure equilibrium.
Periodic boundary conditions are employed throughout, i.e. the upper
$y$-boundary is a second Kelvin-Helmholtz unstable surface.
The resolution is 120$^3$, except for one
run, $\mu$1T5HR, where it is $240^3$. The simulation time is $10^7$
years for each run.

\begin{figure*}[t!]
\resizebox{.47\hsize}{!}{
\includegraphics[clip=true]{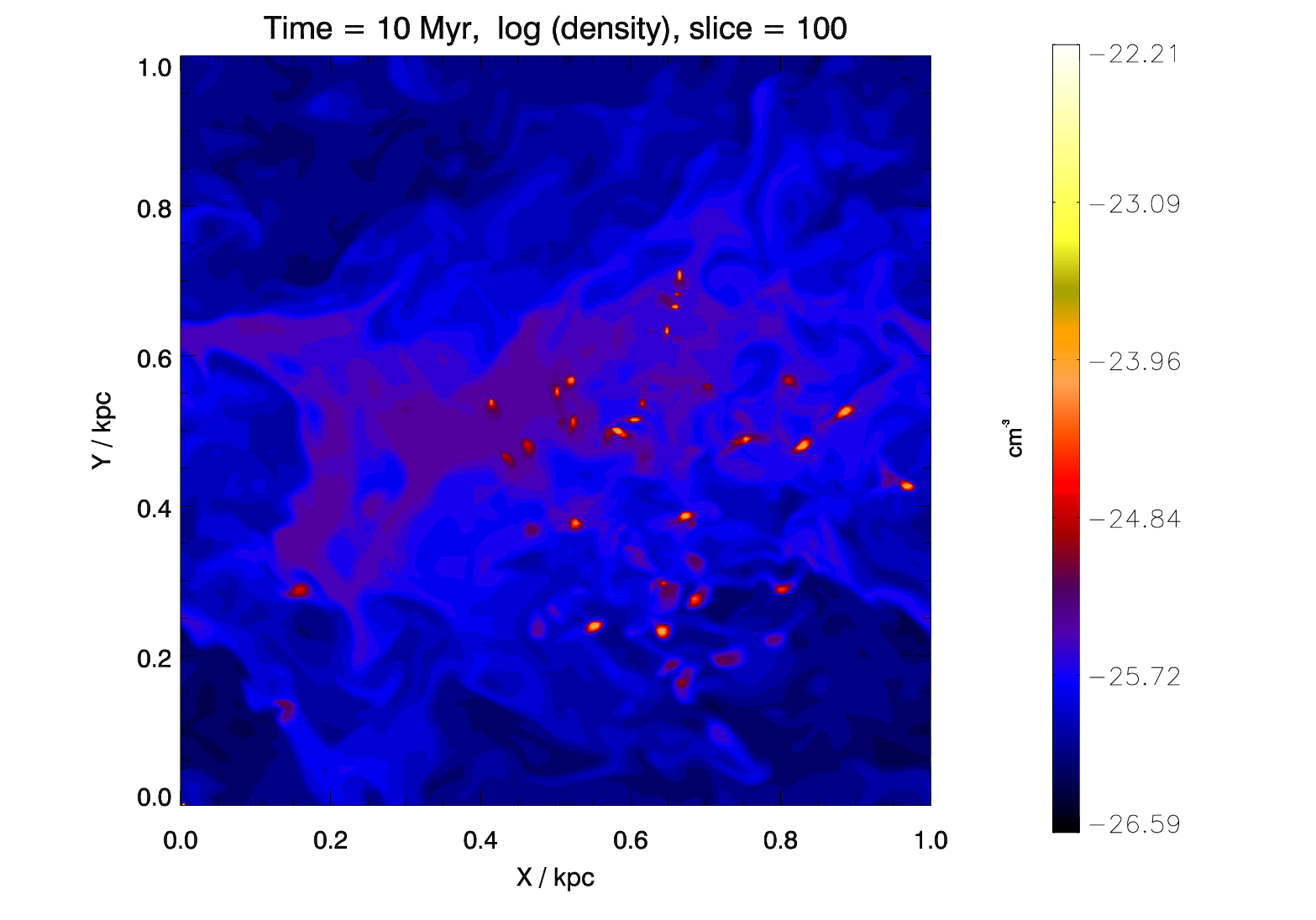}}
\resizebox{.47\hsize}{!}{
\includegraphics[clip=true]{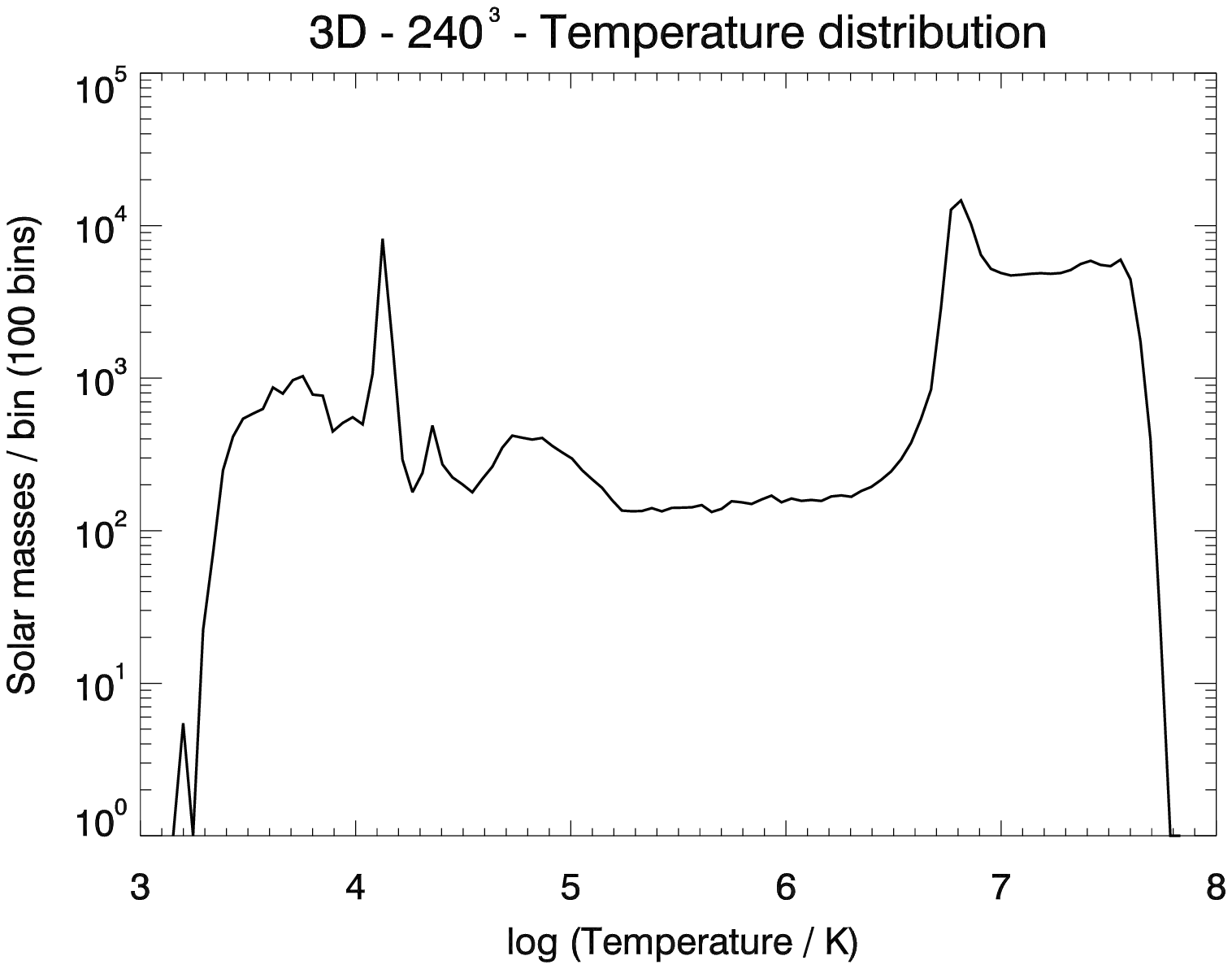}}
\caption{\footnotesize
Left: Density slice through the $\mu$1T5 simulation at 10~Myr. Right:
corresponding temperature histogram.}
\label{plots1}
\end{figure*}

\begin{figure*}[t!]
\resizebox{.47\hsize}{!}{
\includegraphics[clip=true]{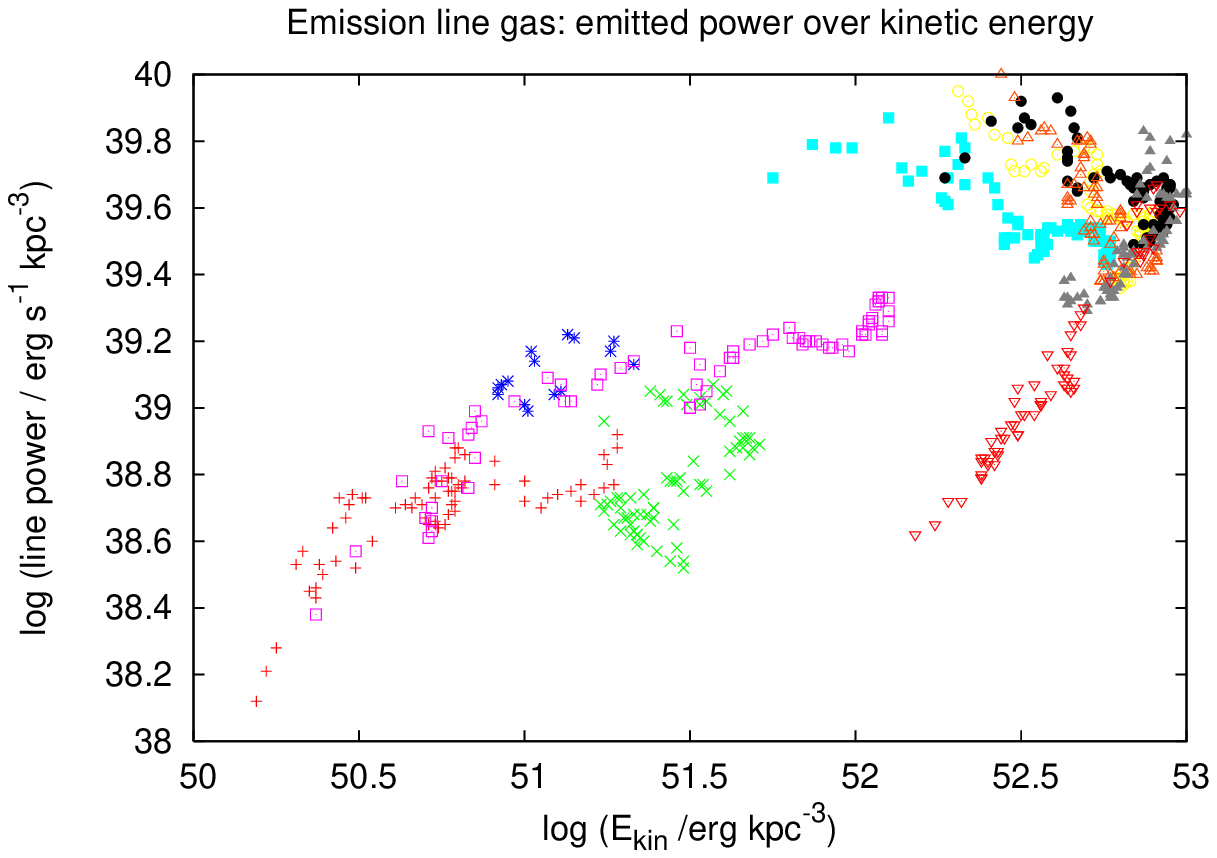}}
\resizebox{.47\hsize}{!}{
\includegraphics[clip=true]{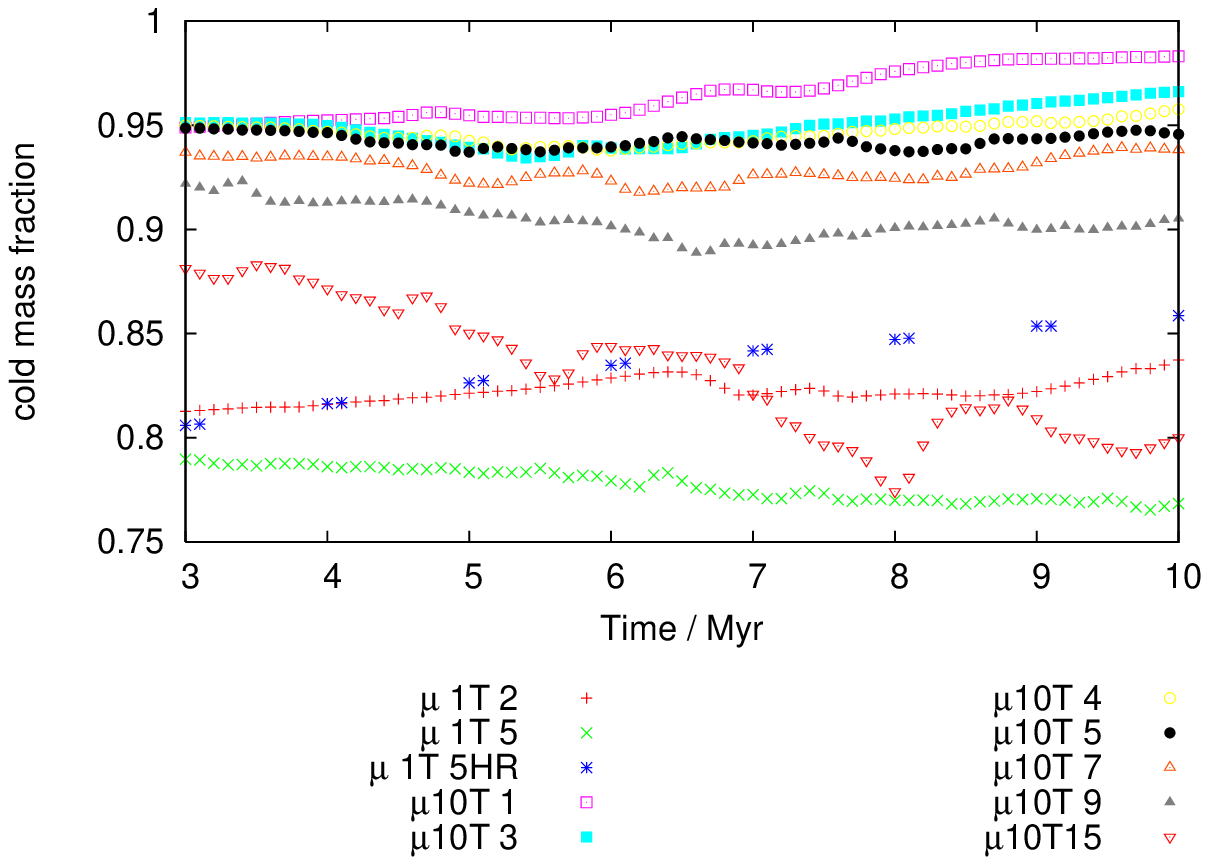}}
\caption{\footnotesize
Left: Radiative power, emitted by the emission line gas (temperature in the
range from 10,000~K to 20,000~K) over Kinetic energy of this
gas. Different symbols are used for different runs shown at many times, 
as indicated in the legend.
$\mu$xTy denotes an initial cloud density of x~m$_\mathrm{p}$~cm$^{-3}$
and an initial temperature of the intermediate component of y~Mio.~K.
Right: Cold gas ($T<10^6$~K) fraction over time. }
\label{plots2}
\end{figure*}

\begin{figure*}[t!]
\resizebox{.47\hsize}{!}{
\includegraphics[clip=true]{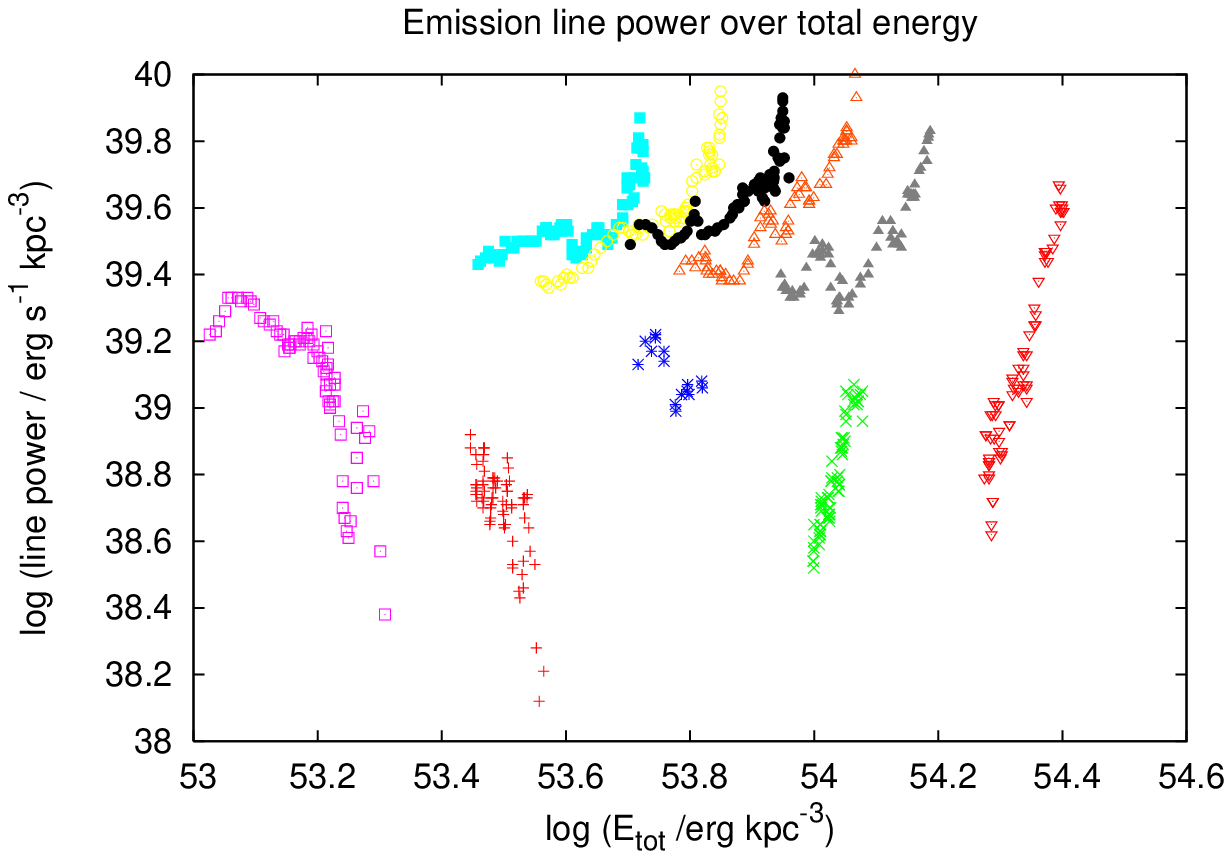}}
\resizebox{.47\hsize}{!}{
\includegraphics[clip=true]{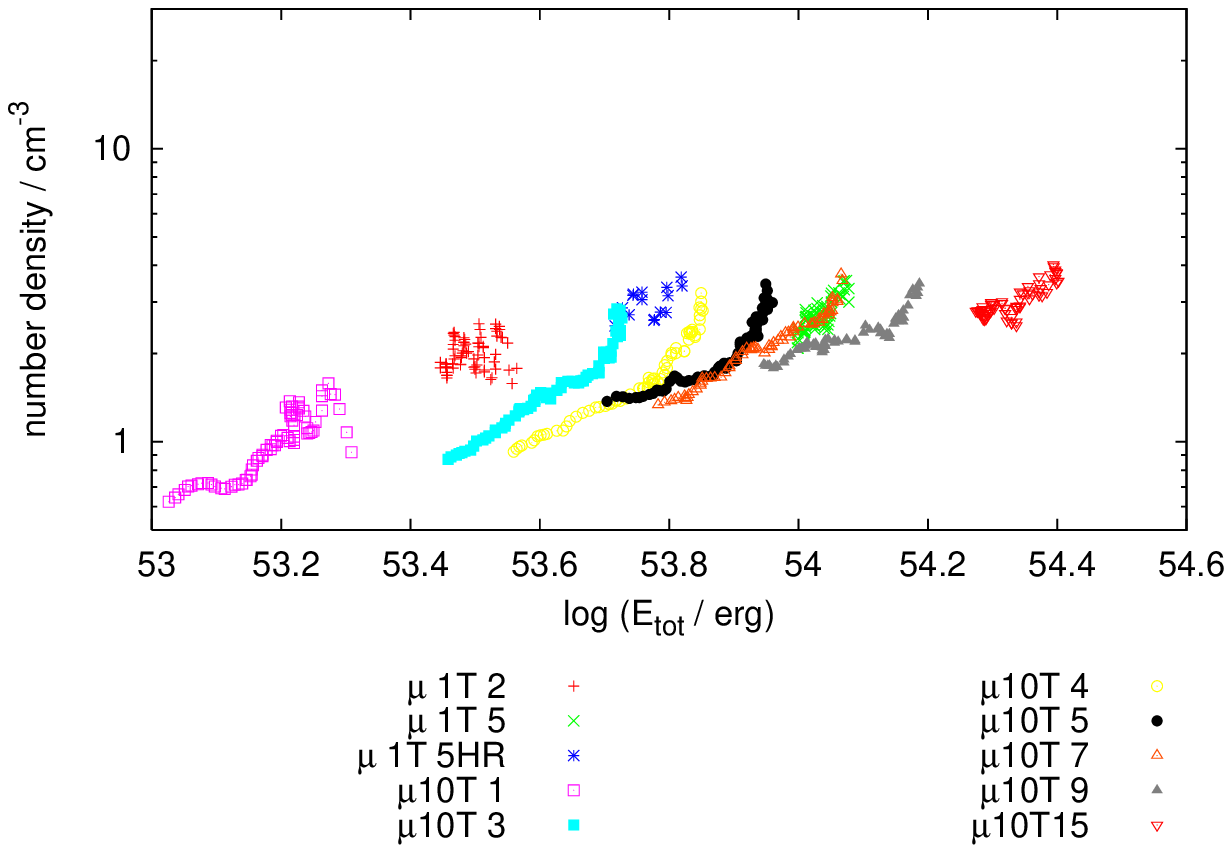}}
\caption{\footnotesize
Left: Emission line power over total energy for different runs 
and simulation times. Right: Number density of the emission line gas 
over total energy.}
\label{plots3}
\end{figure*}
\section{Results}
We show a density slice through run $\mu$1T5HR at the end of the 
simulation in Fig.~\ref{plots1}. The cloud fragments can still be made
out as the densest regions They are often embedded into the
intermediate density gas. The temperature distribution for this
timestep is shown in the same figure. The general trend is 
flatter than in 2D simulations. Still, we see a prominent peak at
14,000~K. There is also a significant accumulation at lower temperature,
and around $10^5$~K. As in our previous 2D simulations, we ascribe the
gas accumulated in the 14,000~K peak, and probably also in the $10^5$
bump to an equilibrium between shock heating and radiative cooling. 
In the following, we call all the gas with temperature in the range
10,000~K to 20,000~K the emission line gas.

An issue for such simulations is weather sufficient resolution was
applied. In principle, the hydrodynamics equations do not include 
a dissipation scale for turbulence, which is artificially set by 
the numerical resolution. Also, the cool clumps would fragment
further, if resolution would allow. However, the processes of interest
may still be adequately represented, as is in fact usually the case 
for turbulence studies: The radiative cooling time of all gas below
about $10^6$~K is much shorter than the simulation time. This gas is
therefore constantly heated. The heating source is either shocks, or 
numerical heat conduction due to bad resolution of contact surfaces.
In the first case, we would expect a correlation of the emission line
power with the ambient temperature, in the second one a correlation
with the kinetic energy. As Fig.~\ref{plots2} shows, the correlation
with the kinetic energy is clearly the dominant one. The correlation,
which most of the simulations follow with a spread of about one Dex, is
linear. About $10^{12}$~ erg are required for the emission of 1~erg/s
in line emission. Interestingly, the two simulations with the highest
external temperature follow a slightly offset pattern, emitting about
a factor of ten less efficiently. The efficiency drop at high
total energy can be seen more clearly in the plots of the emission line
power versus total, and total kinetic energy (Fig.\ref{plots3}). 
This confirms that we really
simulate a realisation of the shock excitation scenario.
The total energy drops monotonically with time, and hence the plots in
Fig.\ref{plots3} May also be understood as an inverse time sequence.
In general, the simulations follow a rather steep line, first, then 
levelling off to a flatter relation. The flatter tracks seem to form 
a kind of equilibrium line, whereas the steeper lines show that the systems
evolve fast towards this equilibrium line.

Also shown in Fig.~\ref{plots2} is the fraction of gas colder than
$10^6$~K against time. There is a dramatic difference between run
$\mu$1T5 and $\mu$1T5HR, which is the same setup at double resolution.
While the low resolution run shows a decline of the cold gas fraction,
it increases in the high resolution one. Yet, for a given resolution,
and initial cold gas fraction, there is a clear trend with the
external temperature in the sense that at low external temperature,
the cold gas fraction drops, whereas it rises at high external
temperature.

\section{Discussion}
While, numerical resolution is an issue here, as always in turbulence
simulations, the line power-kinetic energy correlation clearly shows
that the dominant energy transfer is by shocks. The emitted power
peaks near a  kinetic energy density of $10^{54}$~erg/kpc$^3$.
Interestingly, some
regions in the NLR of M51 \citep{BKB04} are quite close to the 
correlation between the line power and the kinetic energy of the emission 
line gas, none are significantly below, and some are above, as
expected if photo-ionisation comes on top of the shock excitation.
High redshift extended emission line regions have a much higher
kinetic energy density than considered here and emit less efficiently
than indicated by the correlation. This may be hinted at by our
highest energy simulations, but even higher kinetic energy density
simulations are required to address this point.

The typical density in our emission line regions is too low
(Fig.\ref{plots3}), compared
to observations, by a factor of about 100. This may be related to the
quality of these simulations, but could also be a hint that the 
density in the sulfur line emitting regions, which is typically used 
for the density determination \citep[e.g.][]{Nesea06}, may not be typical 
for the majority
of the emission line gas.

An ambient temperature of 5~Mio~K is sufficient for turbulence 
enhanced gas cooling. Given the resolution limitations, higher 
temperatures might still work. Cosmologically, halo temperatures grow
with time, and hence, this gives support to the idea that the external
temperature may be a key factor to explain the absence of extended
NLRs in low redshift radio galaxies.

\bibliographystyle{aa}
\bibliography{references}

\end{document}